\documentclass[prl, aps, twocolumn, superscriptaddress, nofootinbib, tightenlines, nobibnotes, showpacs]{revtex4-1}

\usepackage{slashed}
\usepackage{graphicx}
\usepackage{subfigure}
\usepackage[usenames, dvipsnames]{color}
\usepackage{graphics}
\usepackage{hyperref}
\usepackage{bm}
\usepackage{amsmath}
\usepackage{color}
\usepackage{amsfonts}
\hypersetup{backref,
colorlinks=true,
linkcolor=blue,
linktoc=page,
citecolor=blue,
urlcolor=blue}

\everymath{\displaystyle}

\begin{document}

\title{Axion-plasmon polaritons in strongly magnetized plasmas}

\author{H. Ter\c cas}
\email{hugo.tercas@tecnico.ulisboa.pt}
\affiliation{Instituto de Plasmas e Fus\~ao Nuclear, Lisboa, Portugal}
\affiliation{Instituto Superior T\'ecnico, Lisboa, Portugal}

\author{J. D. Rodrigues}
\affiliation{Instituto de Plasmas e Fus\~ao Nuclear, Lisboa, Portugal}
\affiliation{Instituto Superior T\'ecnico, Lisboa, Portugal}

\author{J. T. Mendon\c ca}
\affiliation{Instituto de Plasmas e Fus\~ao Nuclear, Lisboa, Portugal}
\affiliation{Instituto Superior T\'ecnico, Lisboa, Portugal}


\begin{abstract}

Axions are hypothetical particles related to the violation of the charge-parity symmetry, being the most prone candidates for dark matter. Multiple attempts to prove their existence are currently performed in different physical systems. Here, we anticipate the possibility of the axions coupling to the electrostatic (Langmuir) modes of a strongly magnetized plasma, by showing that a new quasi-particle can be defined, the {\it axion-plasmon polariton}. The excitation of axions can be inferred from the pronounced modification of the dispersion relation of the Langmuir waves, a feature that we estimate to be accessible in state-of-the-art plasma-based experiments. We further show that, under extreme density and magnetic field conditions (e.g. at the interior of dense neutron stars), the axion-plasmon polariton becomes dynamically unstable, similarly to the case of the  Jeans instability occurring in self-gravitating fluids. This latter result anticipates a plausible mechanism to the creation of axion-like particles in the universe. 

\end{abstract}
\maketitle

{\it Introduction.} The violation of the charge-parity (CP) symmetry is perhaps one of the most fundamental problems in modern physics \citep{PhysRevLett.13.138, PhysRevLett.83.22, PhysRevLett.13.138}. Although CP$-$violation is, by construction, inherent to the Standard Model (e.g. it appears as a mixing angle in the Cabibbo-Kobayashi-Maskawa (CKM) matrix describing quark masses \cite{PhysRevLett.10.531, Gell-Mann1960}, and in the Pontecorvo-Maki-Nakagawa-Sakata (PMNS)  matrix mixing lepton flavours \cite{Pontecorvo:1957qd}), there is no evidence of its manifestation in quantum chromodynamics (QCD). At the origin of the so-called {\it strong $CP$ problem} is the anomalous electric dipole moment of the neutron \cite{PhysRevD.92.092003}, which according to a CP$-$broken QCD calculation would be of the order of $10^{-8}$ e.m, while experiments point towards a $10^9$ times larger value. An elegant - and probably the most consensual - way to solve the strong CP problem is the Peccei-Quinn (PQ) mechanism, in which the phase violating the CP$-$ symmetry in the QCD Lagrangian is promoted to a complex field \citep{PhysRevLett.38.1440, RevModPhys.82.557}. The corresponding Goldstone pseudo-boson is known as the {\it axion}, and emerges after breaking the U(1) symmetry in the gluon coupling term \citep{PhysRevLett.40.223}.\par
Axion-like particles (ALPs) are hypothetical particles with a extremely small mass (possibly in the meV range) and couple very weakly with quarks, leptons and photons. ALPs have received renewed breath after being indicated as appealing candidates to dark matter in the universe \cite{PhysRevLett.51.1415, PhysRevLett.104.041301}. Several experiments are in operation for at least a decade with the goal of observing ALP signatures, using both laboratory and astrophysical observations \cite{PhysRevLett.94.121301, PhysRevLett.98.201801,PhysRevLett.118.261301}. Unfortunately, most of the observations are too dubious to confirm the existence of ALPs. For example, the PVLAS experiment - originally design to probe the birefringent properties of the electromagnetic vaccum \cite{PhysRevLett.96.110406} - advanced preliminary results indicating the existence of axions back in 2008, and since then those findings are object of controversial debates (see e.g. \cite{PhysRevLett.98.050402, RevModPhys.82.557,1742-6596-485-1-012035, PhysRevD.84.125019} and referencies therein). \par

The next generation of experiments based on high-power laser facilities is expected to provide unprecedented conditions to probe QED physics in parameter regimes that are inaccessible to particle colliders \cite{RevModPhys.84.1177}. The ELI experiment will offer the possibility to investigate effect of the Heiseinberg-Euler vaccum (virtual electron-hole pairs) \citep{0952-4746-37-1-176, RevModPhys.84.1177} and the quantum recoil due to radiation emission \citep{doi:10.1080/00107514.2014.886840}. The plasma-based wakefield acceleration paradigm gained much breath as it reveals to be an efficient way to accelerate particles \citep{ee183cdbe0634a8dbaabf276f03a1d91, Geddes2004, Faure2004}, and recent studies have exploited such wakefields to produce ALPs in the lab \citep{0295-5075-79-2-21001, 1710.01906, 1751-8121-43-7-075502,1751-8121-49-38-385501}. 

In this Letter, we propose a novel scheme to observe signatures of ALPs in laser-plasma experiments via an axion-plasmon coupling mechanism. In the presence of strong magnetic fields, a new quasi-particle is predicted - the {\it axion-plasmon polariton} -  originating from the hybridization between the axion and the plasma waves. Starting from a PQ-modified electromagnetic theory in the presence of sources, we compute the deformation of the dispersion relation of the electron (Langmuir) waves induced by the axion field. We advance estimates for plasma-based schemes that could demonstrate the existence of axions within the experimentally accessible parameter range. Moreover, we show that axions might be produced inside the interior of neutron stars as a consequence of the dynamical instability of the lower polariton mode, in a mechanism that resembles the Jeans instability occurring in self-gravitating fluids. \par
\begin{figure}[ht!]
\includegraphics[scale=0.67]{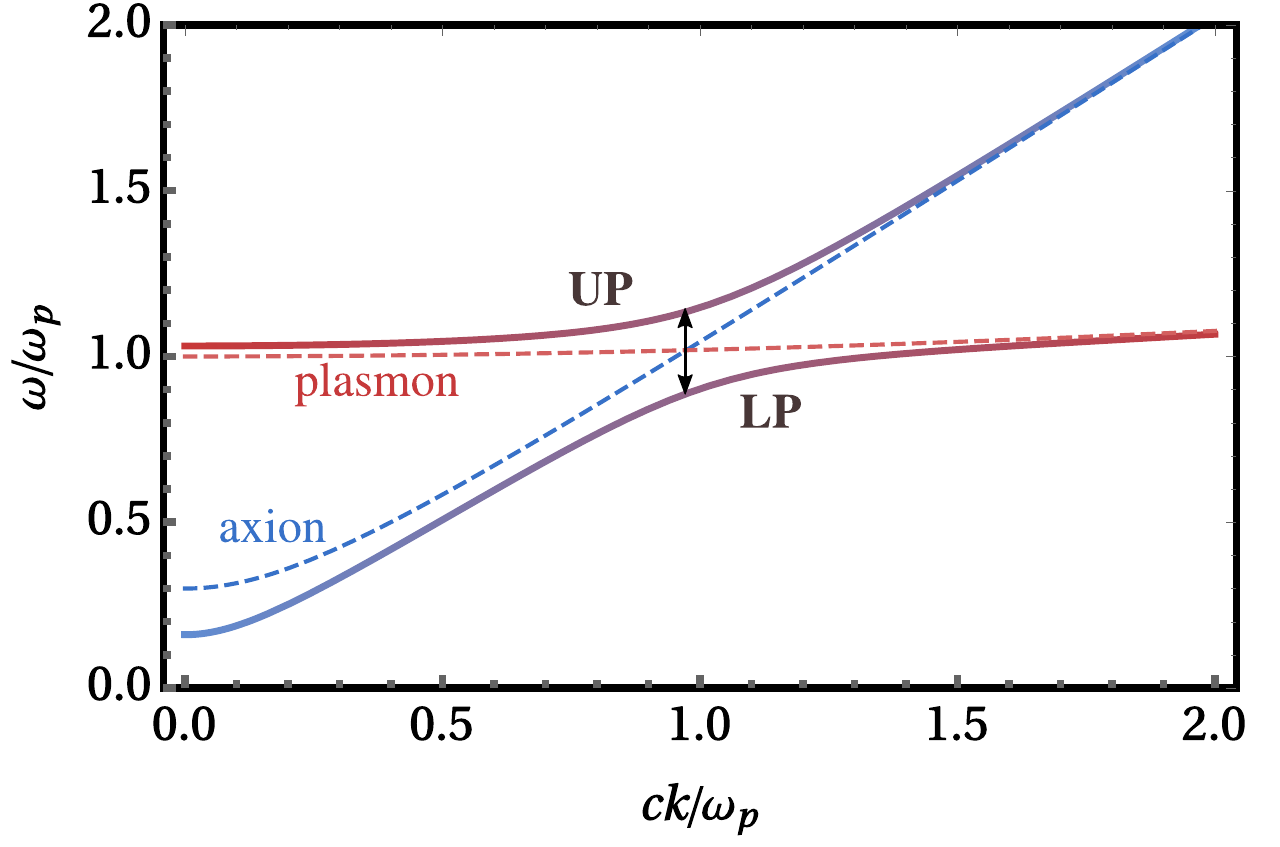}
\caption{(color online) Dispersion relation of the different modes: axion (blue dotted line), plasmon (red dotted line), lower polariton (LU) and upper polariton (UP). At the crossing point $k_*\simeq \omega_p/c$, the LP and UP modes are repelled by $2\Omega$. For illustration, we have set $\Omega=0.5 \omega_p$ and $m_\varphi=0.3 \hbar \omega_p/c^2$.}
\label{fig_dispersion1}
\end{figure}
%
{\it PQ-modified electromagnetism.} 
A minimal electromagnetic theory can be constructed after integrating out the anomalous axion-gluon triangle,  with the effective Lagrangian \citep{doi:10.1142/S0217732313501629}
\begin{equation}
\mathcal{L}=-\frac{1}{4}F_{\mu\nu} F^{\mu \nu}+\mathcal{L}_\varphi+\mathcal{L}_{\rm int},
\end{equation}
where $F_{\mu\nu}=\partial _\mu A_\nu-\partial_\nu A_\mu-A_\mu J^\mu_e$ is the electromagnetic (EM) tensor, $J^\mu_e$ represents the electron four-current, $\mathcal{L}_{\varphi}=\partial_\mu\varphi^*\partial^\mu \varphi/2-m_\varphi^2c^2/(2\hbar^2)\vert \varphi\vert^2$ is the axion Lagrangian and $\varphi$ is the axion field. The axion mass is given by $m_\varphi=\sqrt{z}f_\pi m_\pi/f_\varphi$, where $z=m_u/m_d$ is the ratio between the top and bottom quarks masses, $m_\pi$ is the pion mass and $f_{\varphi(\pi)}$ the axion (pion) decay constant \cite{PhysRevLett.38.1440, PhysRevLett.40.223}. The interaction Lagrangian can be constructed as
\begin{equation}
\mathcal{L}_{\rm int}=\frac{g}{4}\varphi F_{\mu\nu}\tilde F^{\mu\nu},
\end{equation}
where $\tilde F^{\mu\nu}=\epsilon^{\mu\nu\alpha\beta}F_{\alpha \beta}$ is the dual EM tensor, $\epsilon^{\mu\nu\alpha\beta}$ is the Levi-Civita tensor and
\begin{equation}
g=\frac{\alpha_s}{2\pi f_\varphi}\left(\frac{\mathcal{E}}{\mathcal{N}}-\frac{2}{3}\frac{4+z}{1+z}\right)
\end{equation}
is the coupling parameter. Here, $\mathcal{E}$ and $\mathcal{N}$ represent the EM and the color anomalies of the axion current \cite{PhysRevLett.43.103, DINE1981199}. From the Euler-Lagrange equations, we obtain the modified Maxwell equations
\begin{equation}
\begin{array}{c}
\bm \nabla\cdot\left({\bf E}-cg\varphi {\bf B}\right)=\frac{\rho}{\epsilon_0}, \\
\bm \nabla\cdot\left({\bf B}+\frac{g}{c}\varphi {\bf B}\right)=0,\\
\bm \nabla \times \left({\bf E}-cg\varphi {\bf B}\right)=-\frac{\partial }{\partial t}\left({\bf B}+\frac{g}{c}\varphi {\bf E}\right),\\
\bm \nabla \times\left({\bf B}+\frac{g}{c}\varphi {\bf E}\right)=\frac{1}{c^2}\frac{\partial}{\partial t}\left({\bf E}-cg\varphi {\bf B}\right)+\mu_0 {\bf J},
\end{array}
\label{eq_maxwell}
\end{equation}
and the Klein-Gordon equation for the axion field
\begin{equation}
\left(\square +\frac{m_\varphi^2 c^2}{\hbar^2}\right)\varphi=-g{\bf E}\cdot {\bf B}.
\label{eq_KG}
\end{equation}
In a plasma, Eqs. \eqref{eq_maxwell} and \eqref{eq_KG} must be closed with the equations for the sources, $\rho=-e(n_e-n_i)$ and ${\bf u}=-{ \bf J}/(e n_e)$, namely
\begin{equation}
\begin{array}{c}
\frac{\partial n_e }{\partial t}+\bm \nabla \cdot \left(n_e \bf u\right)=0,\\
\left(\frac{\partial }{\partial t}+\bf u\cdot \bm\nabla\right)=-\frac{\bm\nabla P}{m_e n_e}-\frac{e}{m_e}\left(\mathbf{E}+\mathbf{u}\times\mathbf{B}\right),
\end{array}
\label{eq_fluid}
\end{equation}
where $P= k_BT_en_e^\gamma$ is the electron pressure and $\gamma \simeq 3$ is the adiabatic exponent \citep{9781475704617}. In the above equations, we have neglected the ions inertia, as we are interested in the excitation of axions via electron plasma waves only. \par
{\it Axion-plasmon polaritons.} The effect of the axion field in the Langmuir waves can then be determined as follows: consider a strong, homogeneous magnetic field along the $z$-axis, ${\bf B}=B_0 {\bf e}_z$. The electrostatic oscillations along the direction of the magnetic field will then provide the EM energy to excite the axion field, as can be seen from Eq. (\ref{eq_KG}). The created axions will then feedback the plasma via the modified Maxwell equations. Mathematically, this effect can be formulated by assuming fluctuations around the plasma quasi-neutrality condition, $n_e\sim n_0+\tilde n$. Taking the decomposition into Fourier modes, $(\tilde n, \tilde \varphi)\sim e^{ikz-i\omega t}$, and keeping linear terms only, we obtain
\begin{equation}
\begin{array}{c}
\left(\omega^2-\omega_p^2-S_e^2k^2\right)\tilde n-i\frac{gec}{m_e}B_0 k\tilde \varphi=0,\\\\
\left(-\omega^2+\frac{\tilde m_\varphi^2 c^4+c^2k^2}{\hbar^2}\right)\tilde \varphi+i\frac{gec^2}{\epsilon_0 k}B_0\tilde n=0,
\end{array}
\label{eq_modes}
\end{equation}
where $S_e=\sqrt{3k_B T_e/m_e}$ is the plasma thermal speed, $\omega_p=\sqrt{e^2n_0/(\epsilon_0m_e)}$ the plasma frequency and $\tilde m_\varphi=m_\varphi+\hbar g B_0/\sqrt{c}$ the effective axion mass in the plasma. Nontrivial solutions to Eq. (\ref{eq_modes}) implies the secular equation
\begin{equation}
\left(\omega^2-\omega_{\rm pl}^2\right)\left(\omega^2-\omega_\varphi^2\right)-g^2c^3\omega_c^2/(m_e\epsilon_0)=0,
\label{eq_modes1}
\end{equation}
where $\omega_{\rm pl}^2=\omega_p^2+S_e^2k^2$ and $\omega_\varphi^2=\tilde m_\varphi^2 c^4/\hbar^2+c^2k^2$ are the Langmuir and axion bare dispersions, respectively, and $\omega_c=eB_0/m_e$ represents the cyclotron frequency. Solving Eq. \eqref{eq_modes1} yields the lower (L) and upper (U) polariton modes
\begin{equation}
\omega^2_{\rm U, L}=\frac{1}{2}\left(\omega_\varphi^2+\omega_{\rm pl}^2 \pm \sqrt{\left(\omega_{\rm pl}^2-\omega_\varphi^2\right)^2+4\Omega^4}\right),
\label{eq_modes2}
\end{equation}
where $\Omega=(g^2c^3\omega_c^2/\epsilon_0 m_e)^{1/4}$ represents the Rabi frequency. The dispersion \eqref{eq_modes2} describes the hybridization between the axions and the plasmons. If $\Omega$ is larger than the decay rate $\Gamma$ (to be specified below), a new quasiparticle is formed: the axion-plasmon polariton. As such, if axions are excited by the  plasma in the presence of the external magnetic field, then the Langmuir dispersion relation is expected to be strongly modified. In particular, for wave numbers near $\omega_p/c$, the Langmuir dispersion abruptly changes from a flat ($\sim \omega_p$) to a sloped ($\sim c k$) curve. Physically speaking, it means that the upper polariton mode changes from plasmon-like to axion-like near the crossing point $k_*\simeq \omega_p/c$, and conversely for the lower polariton mode. For a typical discharge plasma, $n_0 \sim 10^{10}$ cm$^{-3}$, we estimate the crossing length $\lambda_*=2\pi/k_*\sim 1$ cm, while for a Tokamak plasma, $n_0\sim10^{14}$ cm$^{-3}$, we obtain $\lambda_*\sim 0.1$  mm. This assures the access to the crossing point with simple experimental techniques, such as Langmuir probes. The situation may be slightly different for solid-state target plasmas, $n_0\sim10^{18}$ cm$^{-3}$, for which we have $\lambda_* \sim 1$ $\mu$m, implying the use of more specific techniques. The features of the axion-polariton modes \eqref{eq_modes2} are summarized in Fig. \ref{fig_dispersion1}. \par 

{\it Quantization.} In order to better understand the polariton character of the avoided crossing in Eq. (\ref{eq_modes2}), we proceed to a canonical quantization of the theory. We start by observing that Eq. \eqref{eq_modes1} allows the following decomposition into fast and slow oscillations
\begin{equation}
\begin{array}{c}
\left(\omega^2-\omega_{\rm pl}^2\right)\tilde n=\left(\omega-\omega_{\rm pl}\right)\left(\omega+\omega_{\rm pl}\right)\tilde n\\\\
\left(\omega^2-\omega_{\varphi}^2\right)\tilde \varphi = \left(\omega-\omega_{\varphi}\right)\left(\omega+\omega_{\varphi}\right)\tilde \varphi.
\end{array}
\end{equation}
Since we are interested in the modes that are nearly resonant with the plasma frequency, $\omega\simeq \omega_p$, we perform a rotating-wave approximation (RWA) and inverse Fourier transform Eq. \eqref{eq_modes1} to obtain
\begin{equation}
\begin{array}{c}
\left(i\frac{\partial }{\partial t}-\omega_{\rm pl}\right)\tilde n+i\frac{gec}{2m_e\omega_p}B_0k\tilde{\varphi}=0\\~
\left(i\frac{\partial }{\partial t}-\omega_{\varphi}\right)\tilde \varphi-i\frac{gec^2}{2\epsilon_0\omega_p k}B_0\tilde{n}=0.
\end{array}
\label{eq_RWA}
\end{equation}
We can represent the plasmon and axion field in terms of bosonic operators $\hat a_k$ and $\hat b_k$, obeying the usual commutation relations $\left[\hat{c_k},\hat{c}^\dagger_q\right]=\delta_{k,q}$ ($\hat{c}_k=\{\hat{a}_k,\hat{b}_k\}$), as
\begin{equation}
\tilde n (x)=\sum_k \mathcal{A}e^{ikx} \left(\hat a_k+\hat a_k^\dagger \right), \quad  \tilde \varphi (x)=\sum_k \mathcal{B}e^{ikx} \hat b_k,
\end{equation}
where $\mathcal{A}=k/\sqrt{\omega_{\rm pl}}n_0$ and $\mathcal{B}=c/\sqrt{k\epsilon_0 \omega_\varphi}$ are normalization constants. In terms of these operators, Eq. \eqref{eq_RWA} can be recast as the Heinserberg equations $i \dot {\hat c}_k=[\hat c_k, \hat H]$ associated to the Hamiltonian
\begin{equation}
\hat H= \sum_k\omega_{\rm pl}\hat a_k^\dagger \hat a_k+ \sum_k\omega_{\varphi}\hat b_k^\dagger \hat b_k+\Omega \sum_k\hat{a}_k^\dagger \hat b_k +{\rm h.c.}.
\label{eq_RWA2}
\end{equation}
Full diagnozalization can be obtained by introducing the polariton operators $\hat A_k=u_k\hat a_k-v_k \hat b_k$ and $\hat B_k=v_k \hat b_k+u_k\hat a_k$, yielding
\begin{equation}
\hat H=\sum_k \tilde \omega_L \hat A_k^\dagger \hat A_k+\sum_k \tilde \omega_U \hat A_k^\dagger \hat A_k,
\label{eq_polariton1}
\end{equation}
with $\tilde \omega_{\rm U,L}=\left(\omega_\varphi+\omega_{\rm pl}\pm \sqrt{\Omega^2+(4\omega_\varphi+\omega_{\rm pl})^2}\right)/2$ being the RWA upper (U) and lower (L) polariton modes. Finally, the Hopfield coefficients $u_k$ and $v_k$ satisfy the normalization condition $\vert u_k\vert^2+\vert v_k\vert^2=1$ and read
\begin{equation}
\begin{array}{c}
u_k=\frac{\tilde \omega_{U}\omega_{\rm pl}-\tilde \omega_L\omega_\varphi}{\left(\omega_{\rm pl}+\omega_\varphi\right)\sqrt{\left(\omega_{\rm pl}-\omega_\varphi\right)^2+4\Omega^2}},\\
v_k=\frac{\tilde \omega_{U}\omega_{\varphi}-\tilde \omega_L\omega_{\rm pl}}{\left(\omega_{\rm pl}+\omega_\varphi\right)\sqrt{\left(\omega_{\rm pl}-\omega_\varphi\right)^2+4\Omega^2}}.
\end{array}
\end{equation}
\begin{figure}[t!]
\includegraphics[scale=0.67]{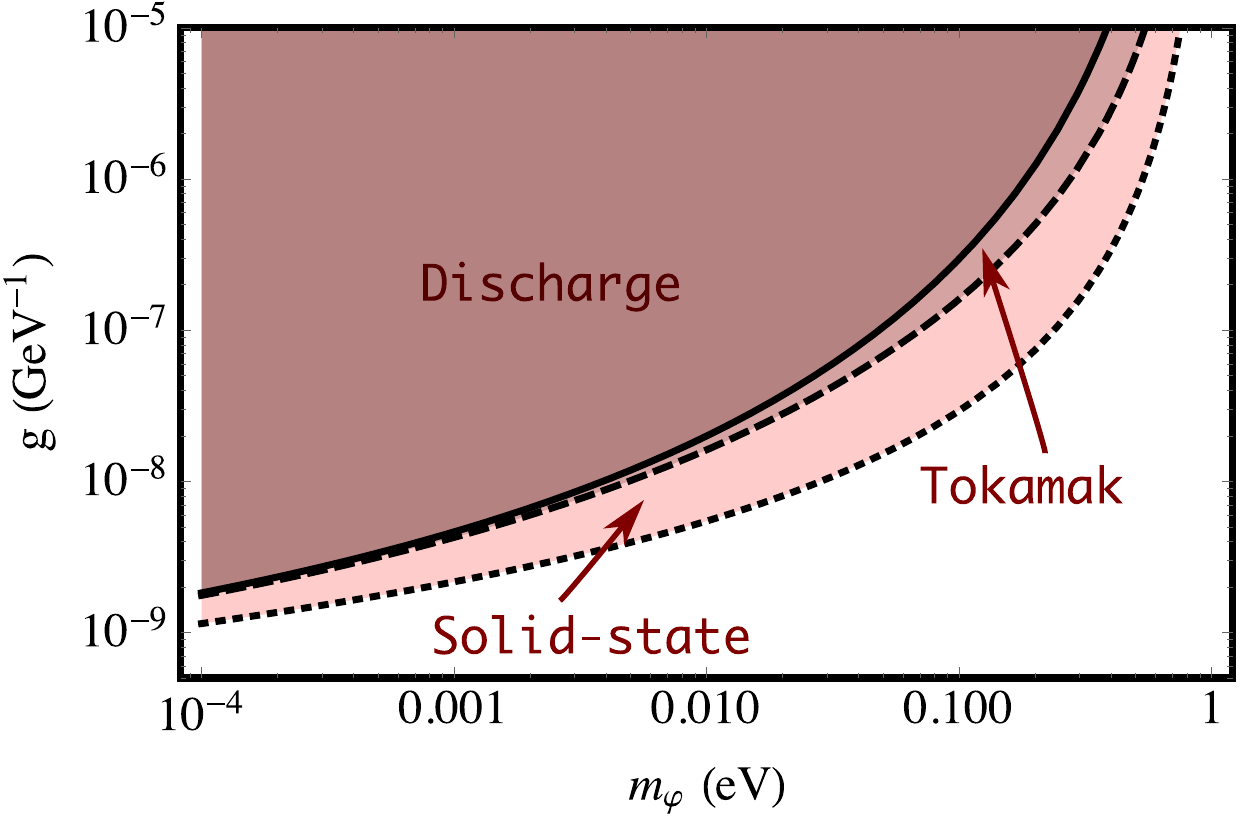}
\caption{(color online) Strong coupling regions (shadowed) as a function of the axion parameters for three different plasmas. Discharge plasmas, $n_0 \sim 10^{10}$ cm$^{-3}$ (solid line), Tokamak plasmas, $n_0 \sim 10^{14}$ cm$^{-3}$ (dot-dashed line), and solid-state target plasmas created by intense lasers, $n_0 \sim 10^{18}$ cm$^{-3}$ (dotted line). In all cases, we have set $B_0=1$ T. }
\label{fig_diagram}
\end{figure}
In oder to obtain Eq. (\ref{eq_polariton1}), we have neglected the counter-rotating terms $\hat a_k^\dagger \hat b_k^\dagger$ and $\hat a_k \hat b_k$, a procedure that is well justified provided the smallness of the coupling parameter $g$ \cite{RevModPhys.82.557}. In fact, by retaining these terms, the diagonalization of Eq. (\ref{eq_RWA2}) yields the full dispersion relation in \eqref{eq_modes2}. For sufficiently strong magnetic fields (e.g. at the interior of extremely dense neutron stars), the inclusion of the counter-rotating terms may be nevertheless necessary, a question that we will address later on. \par
{\it Decay rate.} The second-quantized formalism is particularly helpful to determine the polariton decay rate, a quantity that is crucial to quantify the strength the plasmon-axion coupling. The mode conversion depicted in Fig. \ref{fig_dispersion1} states that, near the crossing point $k_*$, the polariton character abruptly changes from plasma-like to axion-like, and vice-versa. This ``bending" in the dispersion relation is only observable if the decay rate $\Gamma$ is smaller than the Rabi frequency $\Omega$, i.e. provided the strong coupling condition holds. Two decay mechanisms are considered here. The first one is the radiative decay of the axion into two photons, which is given at the rate \cite{RevModPhys.82.557}
\begin{equation}
\Gamma_{\varphi\rightarrow\gamma\gamma}=\frac{g^2 m_\varphi^3}{64\pi}\simeq 1.1 \times 10^{-24}\left(\frac{m_\varphi}{\rm eV}\right)^5~{\rm s}^{-1}.
\label{eq_gamma_phi}
\end{equation}
The second is the decay of axions into plasmons, which can be estimated with the help of Fermi's Golden Rule,
\begin{equation}
\Gamma_{\varphi \rightarrow {\rm pl}} = \frac{2\pi}{\hbar}\sum_{k, q}\vert \mathcal{M}_{k,q}\vert ^2 \delta \left(\hbar \omega_\varphi-\hbar \omega_{\rm pl}\right),
\end{equation}
where $\mathcal{M}_{k,q}=\hbar \Omega \sum_p\langle k\vert  (\hat a_p^\dagger \hat b_p + \hat a_p \hat b_p^\dagger)\vert q \rangle $ is the transition amplitude between the states $\vert k\rangle=\hat a_k^\dagger \vert 0 \rangle$ and $\vert q\rangle=\hat b_q^\dagger \vert 0 \rangle$. Considering only transitions near the crossing point ($\omega_\varphi\simeq \omega_{\rm pl}$), and assuming the axions to be at much lower temperature than the plasmons, we obtain 
\begin{equation}
\Gamma_{\varphi \rightarrow {\rm pl}} \simeq \pi\frac{\Omega^2(1+n_a(\omega_p))^2}{\sqrt{\omega_p^2+m_\varphi^2 c^2/\hbar^2}} ,
\end{equation}
where $n_a(x)=(\exp(x/T)-1)^{-1}$ represents the canonical Bose-Einstein distribution and $T$ is the plasmon temperature. From the Mathiessen rule, the total decay rate is $\Gamma=\Gamma_{\varphi\rightarrow\gamma\gamma}+\Gamma_{\varphi \rightarrow {\rm pl}}$, which is valid since we are here discarding higher-order processes such three-wave mixing (nonlinear decay), which may occur in extremely dense environments or in the presence of shock waves \cite{0741-3335-35-11-009}. In a typical discharge plasma, $n_0\sim 10^{-10}$ cm$^{-3}$, with a magnetic field of $B_0 \sim 1$ T, strong-coupling can be achieved in a region of parameters that has not yet been ruled out by the experiments \cite{Vogel2015}. In Fig. \ref{fig_diagram}, the strong-coupling condition is illustrated as a function of the axion parameters, for different kinds of plasmas. \par
\begin{figure}[t!]
\includegraphics[scale=0.67]{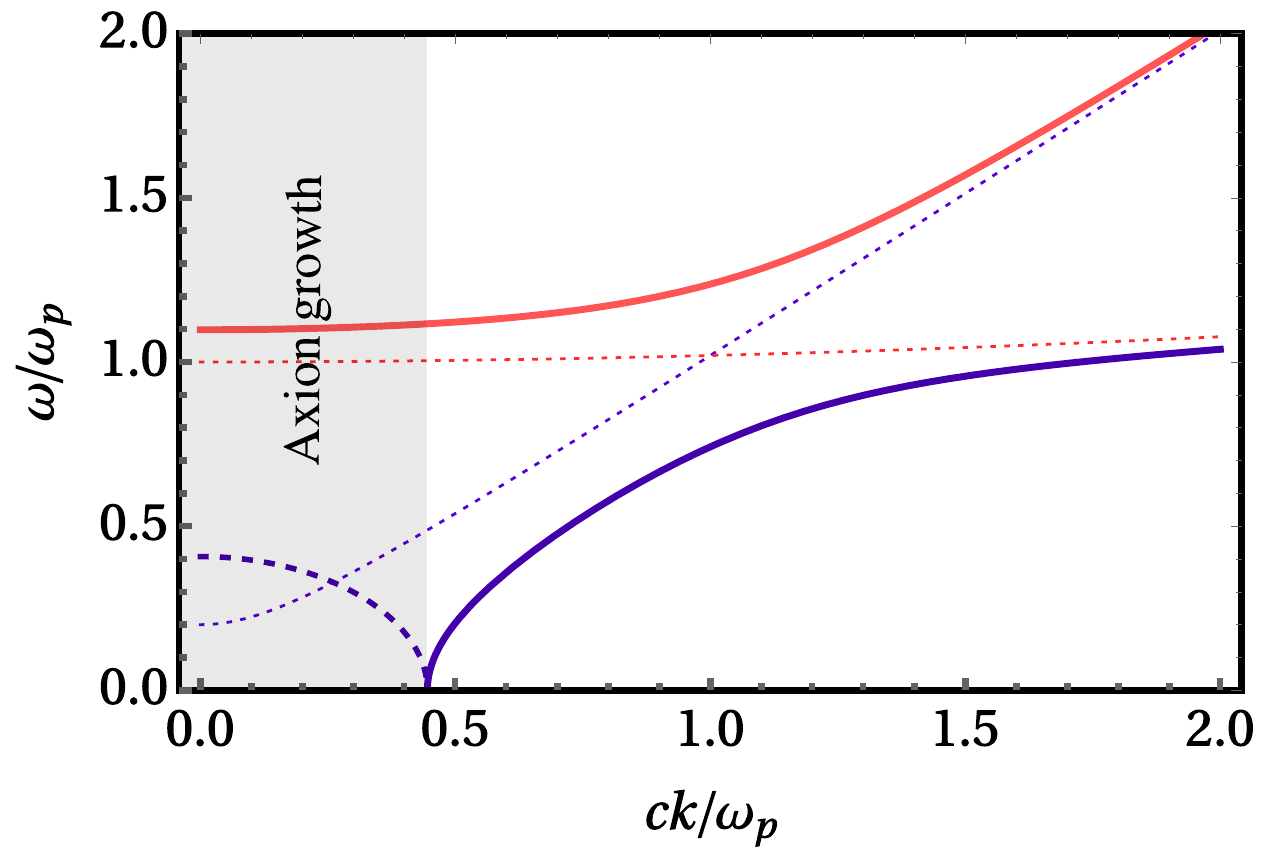}
\caption{(color online) Illustration of the long-wavelength dynamical instability in the lower polariton mode, indicating the growth of axion-like particles in time for $k<k_c$ (see text). The lower solid (dashed) line depicts the real (imaginary) part of $\tilde \omega_U$ in Eq. (\ref{eq_modes2}). We have used the values $g=10^{-9}$ GeV$^{-1}$, $m_\varphi=0.01$ eV, and set $\Omega=0.6 \omega_p$, larger than the instability threshold $\Omega_c$, compatible with extreme conditions at the inner crust of a dense neutron star ($\omega_p=1.5$ MeV and $B_0=10^{10}$ T).}
\label{fig_instability}
\end{figure}

			
{\it Axion instability in neutron stars.} An intriguing situation happens in extremely dense and strongly magnetized plasmas, as it is the case of the interior inner crust of neutron stars. With a typical electron density $n_0\sim 10^{27}$ cm$^{-3}$, plasma frequency of $\omega_p \sim 1.5$ MeV, and magnetic fields up to $B_0\sim 10^{10}$ T, we estimate a Rabi frequency of $\Omega/\omega_p\sim 0.01-10$ (for coupling strengths in the range $10^{-11}-10^{-6}$ GeV$^{-1}$), a quantity that is larger than  the critical value
\begin{equation}
\Omega_c=c\sqrt{\frac{\tilde m_\varphi}{\hbar}\omega_p}\sim (10^{-6}-10^{-3})\omega_p \quad (10^{-4}\leq m_\varphi\leq 1 ~{\rm eV}).
\end{equation}
As such, for $\Omega>\Omega_c$, the wavevectors satisfying the condition
\begin{equation}
k<k_c,\quad {\rm with}\quad k_c=\frac{\sqrt{\Omega^4-\tilde m_\varphi^2c^4\omega_p^2/\hbar^2}}{c\omega_p}
\end{equation}
are dynamically unstable and grow exponentially in time, as depicted in Fig. \ref{fig_instability}. This long-wavelength instability is similar to the Jeans instability in self-gravitating fluids leading to the formation of galaxies \citep{9789401082501, 9780471925675}. In the instability region $k<k_c$, the lower polariton mode is mostly axion-like, i.e. $\vert v_k\vert\gg \vert u_k\vert$, and therefore this suggests that axions may be formed at the interior of neutron stars, as a consequence of the dynamical instability of the lower polariton mode. While a more rigorous estimate of the instability threshold would imply additional details about i) the renormalization of the electron mass and ii) the electron equation-of-state in extreme conditions, the present cold plasma estimates are reasonable at this stage given the actual uncertainty in the axion parameters $g$ and $m_\varphi$.\par

{\it Conclusion.} We have shown that the coupling between the axion and the plasmon in a strongly magnetized plasma may constitute a mechanism to observe the presence of axions. In particular, the most important feature is the avoided crossing between the Langmuir and the axionic Klein-Gordon dispersions near the resonance frequency. The consequence of such a mode repulsion is the changing in character of the Langmuir mode, which changes from electrostatic ($\omega\sim \omega_p$) to an electromagnetic one ($\omega\sim ck$). The strong coupling condition (i.e. the criterion for the axion-plasmon polariton to be a well-defined quasiparticle) has been calculated for different types of plasmas. Additionally, in extreme situations, such at the interior of neutron stars, the combination of strong magnetic fields and extremely high electronic densities is expected to lead to a dynamical instability in the lower polariton mode, which is mostly axionic in character. This suggests that axion-like particles may be produced as a consequence of an instability mechanism, similarly to the Jeans instability in self-gravitating fluids leading for the formation of galaxies. 
\par
HT acknowledges FCT - Funda\c{c}\~{a}o da Ci\^{e}ncia e Tecnologia (Portugal) through the grant number IF/00433/2015. JDR acknowledges the Doctoral Programme in Physics and Mathematics of Information (DP-PMI) and the financial support of FCT - Funda\c{c}\~{a}o da Ci\^{e}ncia e Tecnologia through the grant number SFRH/BD/52323/2013.

\bibliographystyle[maxnames=5]{apsrev4-1}
\bibliography{axion_references}
\bibliographystyle{apsrev4-1}

\end{document}